\documentclass[prd,showpacs,preprint,amsfonts]{revtex4}
\usepackage{epsfig}
\usepackage{graphics}
\usepackage{amsmath}

\newcommand{\Eq}[1]{Eq.~(\ref{#1})}
\newcommand{\reff}[1]{(\ref{#1})}
\newcommand{\beq}{\begin{equation}}
\newcommand{\eeq}{\end{equation}}
\newcommand{\bea}{\begin{eqnarray}}
\newcommand{\eea}{\end{eqnarray}}
\def\half{{\scriptstyle{\frac{1}{2}}}}

\begin{document}

\author{Christopher Eling and Jacob D. Bekenstein}
\affiliation{Racah Institute of Physics, Hebrew University of
Jerusalem, Jerusalem 91904, Israel\\}\date{\today}
\title{Challenging the generalized second law}
\begin{abstract}

The generalized second law (GSL) of black hole thermodynamics states
that the sum of changes in black hole entropy and the ordinary
entropy of matter and fields outside the hole must be non-negative.
In the classical limit, the GSL reduces to Hawking's area theorem.
Neither law identifies the specific effects which makes it work in
particular situations. Motivated by Davies' recent \textit{gedanken}
experiment he used to infer a bound on the size of the fine structure
constant from the GSL, we study a series of variants in which an
electric test charge is lowered to a finite radius and then dropped
into a Schwarzschild, a near-extremal magnetic Reissner-Nordstr\" om
or a near-extremal Kerr black hole. For a classical charge, we
demonstrate that a specific ``backreaction" effect is responsible
for protecting the area theorem in the near-extremal examples. For
the magnetically charged Reissner-Nordstr\" om hole an area theorem
violation is defused by taking into account a subtle source of
repulsion of the charge: the spinning up of the black hole in the
process of bringing the charge down to its dropping point. In Kerr
hole case, the electric self-force on the charge is sufficient to
right matters. However, in all experiments involving an elementary
charge, the full GSL would apparently be violated were the fine
structure constant greater than about order unity. We argue that in
this case a quantum effect, the Unruh-Wald quantum buoyancy, may
protect the GSL.

\end{abstract}
\maketitle

\section{Introduction}
\label{sec:intro}

One of the most useful general results in black hole physics is
Hawking's area theorem~\cite{Hawking:area}: in the presence of
matter and fields which obey the weak positive energy condition, the
area of the event horizon of any black hole cannot decrease.  This
theorem is proved under the assumption that the congruence of the
horizon's generators has no caustics in the future, in other words,
that the generators do not run into a future singularity.  The
theorem does not identify specific effects which make it work in
particular situations.  A venerable application of this theorem is
to demonstrate the necessity of superradiance by charged or rotating
black holes~\cite{Bek:super}.  The theorem is also closely related
to black hole thermodynamics: the generalized second law (GSL) for
black holes~\cite{Bek:GSL} is an extension of the area theorem which
reduces back to it when Hawking's radiation has negligible effect.

Recently, Davies~\cite{Davies:2007wa} has employed the GSL (and thus
thermodynamics) to establish an upper bound on the fine structure
constant, and to infer that a magnetic monopoles's spatial extent
considerably exceeds its Compton length.  He envisages a
\textit{gedanken} experiment in which a charged point particle is
dropped from rest at infinity into a Schwarzschild black hole.  By
assuming that the GSL reduces to the area theorem, and consequently
demanding that the area of the horizon grow upon ingestion of the
charge, Davies concludes that the fine structure constant cannot be
larger than unity.  This derivation is interesting because of its
thermodynamic origin.  It immediately raises the question whether
variations on the Davies \textit{gedanken} experiment can yield
additional insights into the possible bound and further conclusions
of physical interest. To answer this question we examined a comprehensive
set of variants of Davies' designed to ``push 
the area theorem against the wall'': the dropping of either an extended or point charge \textit{from rest at a finite distance} from a Schwarzschild, Reissner-Nordstr\" om
(RN), and Kerr hole.

We start in Sec.~\ref{sec:constraints} by reviewing Davies'
argument, and showing that the key assumption is that the dropped particle is
point-like. Davies also assumes the Hawking emission is
insignificant, but we point out that in the regime where the bound
is obtained, the  Hawking radiation entropy  cannot actually 
be neglected in the GSL. To characterize this contribution we
modify the experiment, imagining instead a series of point of
charges are dropped into the hole. The result is a slightly weakened
upper bound still close to Davies' value, provided we are allowed to
make the timescale of the dropping appropriately small.

With this \textit{gedanken} experiment in mind,  we consider our new
lowering--dropping process for  a trio of black hole cases.
First, in Section \ref{sec:lowering}, we allow the dropped object to
be an extended classical charge.   We start (Sec.\ref{sec:setup}) by enumerating the
conditions that the object's charge, mass and radius must obey in
relation to the hole's mass in order that various complications may
be avoided. We then consider (Sec.~\ref{Schwarzschild})  the
dropping of the object from rest at a finite distance into a
Schwarzschild black hole. We point out that one should take into
account, not only the gravitational force, but also the repulsive
electromagnetic self-force the object is subject to. Analytic
arguments show that no violation of the area theorem ensues in this
case.

To further challenge the theorem we replace the Schwarzschild black hole by an
almost extreme magnetically charged RN one
(Sec.~\ref{sec:monopole}). Here a violation of the area theorem
would occur for classical charged objects satisfying the weak energy
condition, even with the self-force accounted for.  But, as
made clear numerically, the theorem is saved by the intervention of
an extra repulsion.   The electric dipole moment induced in the hole
by the angular momentum gradually transferred to it repels the
charge as it is lowered. As a final example we look
(Sec.~\ref{Kerr}) at the case of a charged object lowered along the
symmetry axis of a nearly extreme (neutral) Kerr black hole down to
a certain point, and then dropped into it. Here the only known force
acting on the charge, apart from gravity, is the self-force. We
compute this force anew, and show that if it is not properly
accounted for, another inconsistency arises between the area theorem
and the weak energy condition.

In Sec.~\ref{ep} we consider lowering and dropping an elementary
charge into our black holes. We allow for the classical
``backreaction" effects discussed in the previous section, and employ a large black hole to minimize the effect of the Hawking emission. We
find that the GSL, which  reduces to the area theorem in this case,
seems to imply an upper bound on the fine structure constant $\alpha
\lesssim 2$, similar to Davies' bound. Based on these results, must
one conclude that black hole thermodynamics truly requires a bound
on the magnitude of the electromagnetic coupling in nature? We argue
that in this case an entirely \textit{quantum} effect, the average
repulsive force on the elementary charge due to Compton scattering
of photons from the black hole's ``thermal atmosphere", an example
of Unruh-Wald buoyancy, should not be neglected, and may be the
protector of the GSL.

Sec.~\ref{discussion} summarizes our results and discusses possible
future work. We also note that in the \textit{gedanken} experiments
where charges are dropped from infinity, strong self-field effects,
which were neglected previously, are not negligible and could cause
the process to break down. Throughout this paper we use units
with $c=G=1$, but display $\hbar$; hence $\surd\hbar=m_P$, the Planck mass. Numerical values and electromagnetic relations are stated with electrostatic units in mind.

\section{Constraints on fine structure constant from the GSL?}
\label{sec:constraints}

\subsection{Davies' argument}

We first review the argument of Davies~\cite{Davies:2007wa} for setting
constraints on the value of the fine structure constant and on the
physical size of magnetic monopoles from the GSL. Davies considered
 a simple gedanken experiment where a \textit{point} particle of electric (or
magnetic) charge $g$ with mass $m$ is dropped from rest at infinity into a
Schwarzschild black hole of mass of $M$. It is assumed that $g, m
\ll M$ so that the in-falling test particle moves along the
geodesics of the background Schwarzschild geometry. When the
particle is absorbed, the black hole becomes an electric or magnetic
RN solution. Therefore the change in the horizon
cross-sectional area in the process is
\beq
\label{dA}
 \delta A = 4\pi(M+m+\sqrt{(M+m)^2-g^2})^2-16\pi M^2. \eeq
Obviously $\delta A< 0$ whenever  $ M+m+\sqrt{(M+m)^2-g^2}< 2M$,
from which follows the \textit{exact} criterion for the area to decrease,
and for the the GSL (more correctly the area
theorem) to be violated:
\beq g^2/m > 4M. \label{Daviesschw}\eeq

At this point Davies invokes quantum mechanics:  the effective
``radius'' of the particle cannot be smaller than approximately its
own Compton wavelength $\lambda_{\rm C}= \hbar/m$; hence the black hole
diameter $4M$ must be greater than the last in order for the
particle to be absorbed by, rather than scatter off the black hole.
Therefore, the perceived violation of the area theorem entails
\beq g^2/\hbar \gtrsim 1. \label{Daviesh}
 \eeq
Accordingly, in Davies' view, the GSL requires that in nature
\beq
g^2/\hbar\lesssim 1.
\label{Daviesh2}
\eeq

Should we be incredulous that such fundamental bound on a coupling
constant results from thermodynamics? With the electron's charge $e$
in  mind, Davies reminds us that  a value of the  fine structure
constant as large  as $ \alpha = e^2/\hbar \gtrsim 1$ would induce
spontaneous electron-positron pair creation in the vacuum~~\cite{Zeldovich}, and
is thus unviable.

Whenever $g$ represents a magnetic monopole charge $\mu$,
the Dirac quantization condition leads to the basic constraint
\beq \mu e \geq \hbar/2. \label{Dirac}\eeq
Now in terms of the observed fine structure constant $\alpha \approx
1/137$, this inequality becomes $\mu^2/\hbar \gtrsim (137/4)$, which
clearly contradicts the aforementioned criterion
$\mu^2/\hbar\lesssim 1$ for protecting the GSL. To circumvent this
problem, Davies proposes that magnetic monopoles are never point
particles, but rather extended objects with some effective proper
radius $R\gg \hbar/m$. Since $R \lesssim 2M$ for the particle to be
absorbed by the hole, he arrives at the new criterion,
\beq \mu^2 \lesssim m R, \label{monopole}\eeq
for the area theorem to hold. Combining this with (\ref{Dirac}) yields
the additional constraint
\beq {e^2\over \hbar} \gtrsim {\hbar/m\over R}. \label{alphamono}\eeq
on the size of the fine structure constants in a world
with monopoles.

\subsection{Critique}\label{sec:crit}

Davies assumes that the particle being dropped is a point particle,
i.e., one whose localization scale is its own Compton length
$\lambda_{\rm C}=\hbar/m$.  He further tacitly assumes that the black hole
is so massive that production of entropy via the Hawking
radiation~\cite{Hawking:rad} is completely negligible compared to
the black hole entropy change in the process contemplated here.  To
see that these two assumptions are indispensable for his conclusions, let
us imagine a purely classical variant of the argument.

Classically there is no GSL, but only the Hawking area theorem. One
should not need to invoke quantum mechanics to prevent the theorem's
violation when inequality (\ref{Daviesschw}) holds.  Let us recall
that $g^2/m$ constitutes the ``classical radius''  $r_c$ of our
charged object or particle, the radius the object (regarded as
spherical) would have  were its rest mass $m$ to be comprised
exclusively of electromagnetic energy. Therefore, if we think of our
charge as an extended classical object, its effective proper radius
$R$ must exceed $r_c$ in order for the object's energy density to be
positive everywhere (for an electron in the real
world $r_c \approx \lambda_{\rm C}/137$). If \Eq{Daviesschw} holds so that
$\delta A < 0$, and also $R \lesssim M$ so the object can indeed be
absorbed, we find that $r_c \gtrsim R$. But this just means that the
object contains regions with negative energy density. Thus  the
violation of the area theorem is understandable: the theorem is
always proved under the assumption that the weak positive energy
condition is respected.   By working entirely within classical
physics we fail to come up with significant conclusions.

Let us now return to the quantum elementary particle.  To be
definite we take $g$ to be the electric charge, $e$. We note that
$\lambda_{\rm C}= r_c/\alpha$.  Thus in a world where $\alpha>1$,
$\lambda_{\rm C}< r_c$, and we can well ask, what is the actual size of
our charge, $\lambda_{\rm C}$ as would be suggested by quantum
considerations, or $r_c$ as would be required by positivity of the
internal energy density?  In the former case the condition for the
charge to be absorbed by the black hole is $\lambda_{\rm C}<4M$ and in the
latter $r_c<4M$.   In the first case we can rework the condition
$\alpha>1$ into the form $e\cdot (e/\lambda_{\rm C}{}^2)\lambda_{\rm C} > m$
which says that the electric field of our charge can impart to a
like (virtual) particle in the vacuum an energy exceeding its own
rest mass while that particle is a Compton length or so away.  In
the second case $\alpha>1$ is equivalent to $e\cdot (e/r_c{}^2)r_c>
m$ so that, again, our charge can impart to a like virtual charge in
the vacuum an energy greater than its rest mass while the virtual
charge is a distance $r_c$ away.

Thus under both assumptions about the size of our charge, we would
expect it to strongly polarize (if not outrightly break down) the
vacuum adjacent to it. But then the expectation value of the
stress-energy operator $\hat{T}_{\mu \nu}$, which appears as the
source in the semiclassical Einstein equation,
\beq G_{\mu \nu} = 8 \pi <\hat{T}_{\mu \nu}>, \eeq
need no longer satisfy the weak energy condition in the particle's
neighborhood. This weakens the basis for the validity  the area
theorem in the situation we have in mind. Recall that in the Hawking
radiation process, vacuum polarization opens a quantum loophole in
the area theorem, and this is just what allows the black hole to
evaporate and decrease its horizon area.  Furthermore, in the first
case the charge is smaller than its classical radius, which means,
classically speaking, that its interior has negative energy density
somewhere.  For all these reasons one cannot, \textit{a priori},
rely on the validity of the area theorem when $\alpha>1$.

However, one can appeal to the presumably more reliable GSL,
\beq \delta S_{\rm ext} + {\scriptstyle 1\over\scriptstyle 4}\hbar^{-1} \delta A \geq 0,
\label{eq:GSL}\eeq
namely, that if the  black hole area decreases, the decrease must be
more than compensated by generation of a suitable amount of
radiation entropy.  Davies tacitly assumes that the term  $\delta
S_{\rm ext}$ can be neglected here, so that the GSL reduces to the
area theorem. One might think this would be so for a sufficiently
massive black hole, in which case Hawking radiance would be
suppressed.  Let us check.  By expanding \Eq{dA}  for $e,m\ll M$
we obtain the black hole entropy change contributed by the
particle's in-fall alone:
\begin{equation}
{\scriptstyle 1\over\scriptstyle 4}\hbar^{-1} \delta A\approx
(2\pi/\hbar)(4Mm-e^2).
\end{equation}

During the particle's disappearance, the hole radiates.
Approximating this as emission of thermal radiation from a sphere of
radius $2M$ at temperature $\hbar(8\pi M)^{-1}$, we find that the
rate of thermal entropy emission is $\dot S_{\rm rad}\approx (1920
M)^{-1}$.   This is a contribution to $ \delta S_{\rm ext}$. The
emission induces a decrease in black hole entropy which scales with
$M$ just as does the radiation entropy, and whose rate is known to be
somewhat smaller in magnitude than $\dot S_{\rm rad}$; it makes a
contribution to $\delta A$.  Both together amount to
\begin{equation}
\dot S_{\rm BH}+\dot S_{\rm rad}= \gamma(1920 M)^{-1}
\label{Hawkcont},
\end{equation}
where the factor $\gamma$ is  dimensionless, positive and of order unity.
Putting together all above contributions we may write \Eq{eq:GSL} as
\begin{equation}
\frac{\gamma\Delta t}{1920 M}+\frac{2\pi(4Mm-e^2)}{\hbar}\geq 0,
\label{GSL2}
\end{equation}
where $\Delta t$ is the global time that elapses as the charge goes down  the hole.

What to take for $\Delta t$? For one charge it is a somewhat
ambiguous quantity. So it helps to consider a series of like charges
being dropped sequentially from rest at global time intervals
$\Delta t$, with their starting points distributed around a sphere
at large distance from the black hole. By the stationary character
of Schwarzschild's spacetime, the charges will arrive at a specified
radius $r$ near the horizon also separated by time intervals $\Delta
t$.  Davies' bound comes from assuming that the charge is spatially
almost as big as the black hole. Therefore when one charge is just
beginning to be absorbed by the black hole (i.e. its ``center of
mass" is about proper distance $R$ from the horizon), its spatial
separation from the subsequent charge must be at least $R$. This is
the minimum criterion for the charges to avoid bumping into one
another. In the Schwarzschild metric the rate of change of the
proper distance from the horizon $\ell$ for a freely falling
particle is
\begin{equation}
d\ell/dt=-\sqrt{(2M/r)(1-2M/r)}.
\end{equation}
This formula can be used to calculate the time separation $\Delta
t$. However, since $r$ should be taken somewhat larger than $2M$, we
can simply parameterize $\Delta t$ as
\begin{equation}
\Delta t=\kappa R.
\end{equation}
with $\kappa$ at least a few times unity.  We think of $\kappa$ as fixed.

In our first case ($R=\hbar/m$), we have $\Delta t=\kappa \hbar/m$, so that \Eq{GSL2} gives
\begin{equation}
\alpha\leq \frac{4Mm}{\hbar}+\frac{\gamma\kappa\hbar}{3840\pi Mm}.
\label{newbound}\end{equation}
It is immediately clear that large $M$ is not most propitious for
deriving the tightest bound on $\alpha$. Let us instead minimize the
r.h.s. of \Eq{newbound} with respect to $M$ while respecting Davies'
requirement that $\hbar/m<4M$.  We get
\begin{equation}
\alpha\leq \max\left(1+\frac{\gamma\kappa}{960\pi},\sqrt\frac{\gamma\kappa}{240\pi}\,\right).
\end{equation}
\textit{If} indeed $\kappa$ can be made just a few times unity, the
upper bound on $\alpha$ is quite close to unity, as Davies contends.
In the second case ($R=e^2/m=\alpha\hbar/m$), the appropriate
parametrization is $\Delta t=\kappa\alpha\hbar/m$.  This will
transform \Eq{GSL2} into
\begin{equation}
\left(1-\frac{\gamma\kappa\hbar}{3840\pi Mm}  \right)\alpha\leq \frac{4Mm}{\hbar} .
\label{newbound2}
\end{equation}
The additional demand that $e^2/m<4M$, or $4Mm/\hbar>\alpha$,
automatically causes \Eq{newbound2} to be satisfied for all values
of $\kappa$.  So in this case the GSL sets no bound whatsoever on
$\alpha$.

We conclude that if, in a world with $\alpha>1$, an elementary
charge's  spatial extent is its Compton length, then Davies' bound
on $\alpha$, or one very close to it, can indeed be derived from the
GSL.  This is contingent on existence of an appropriate charge
dropping procedure with a $\kappa$ not very large on scale unity, one without instabilities of the train of charges due to their Coulomb repulsion. It is also contingent on resolution of a quandary to be raised in Sec.~\ref{discussion}.  If when $\alpha>1$ an elementary charge is as large as its classical radius, no bound son $\alpha$ can be derived from the GSL.

\section{Lowering a classical charge and then dropping it}
\label{sec:lowering}

Given the above caveats on Davies' argument, can variants of
the Davies' \textit{gedanken} experiment yield stronger constraints on $\alpha$, or even new conclusions?  We first explore a variant which focuses on a
classical charged object which is dropped from the vicinity of the
black hole instead of from infinite distance. The new scenario is
advantageous since the conserved Killing energy of the object, when
ultimately dropped, is smaller than that pertaining to the object
dropped from rest at spatial infinity.  The consequent reduction of
$\delta A$ is more likely to challenge the area theorem.
Furthermore, the use of a classical object allows us to escape the
complications of the analysis of Sec.~\ref{sec:crit}. Given that
Davies' bound emerges when the black hole radius is almost as small
as the elementary  particle's, it would make little sense there to
ignore Hawking's radiance, which is the more intense the smaller the
black hole.  Thus since vacuum polarization issues force us to trade
area theorem for GSL, we cannot avoid dealing with complications due
to the radiation.  By switching to a classical charged object, we no
longer need to consider black holes with microscopic radius, and so
we may hope the radiation plays a negligible role. We also explore
whether use of RN or Kerr black holes, instead of Schwarzschild
ones, can expose new things. An allied question we ask is whether
there are any other forces, apart from gravitation and Coulomb ones,
which are crucial for the outcome of the \textit{gedanken}
experiment.

\subsection{Setup}
\label{sec:setup}

We imagine that a classical object with charge $q$ is first brought
to rest at some finite radius from the horizon of a black hole,
which we take of Schwarzschild, magnetic RN, or Kerr type, and then
dropped freely.  How is the charged object deposited at a finite
radius? It could be lowered adiabatically from infinity with an
appropriately designed apparatus. One possibility for this,
discussed for the Schwarzschild case in
Ref.~\onlinecite{Fouxon:2008pz}, is a conical cable consisting of
radial fibers filling the portion of space defined by some solid
angle with vertex at the black hole. This cable can be modeled by
the stress tensor $T^\mu{}_{\nu} = {\rm diag}[-\rho(r),S(r),0,0]$,
where $\rho(r)$ is the proper density and $S(r)$ the tension. It
turns out that this stress tensor respects the weak energy condition
$|S|/\rho \leq 1$; therefore, the cable need not snap under tension
at a finite radius from the hole~\cite{Fouxon:2008pz} (see also our
Appendix~B for a demonstration that a properly constructed cable also will not snap in the Kerr spacetime). We consider the adiabatic
lowering process to be on a time scale $\tau$ much longer than the
inverse surface gravity $\kappa^{-1}$ of the black hole. Thus the
change in horizon area in the course of the lowering process will be
negligible~\cite{membrane}.

In what follows we treat the object as a uniformly charged sphere of
proper radius $R$. We compute in the test particle approximation. In
addition to the obvious requirement that ${m,q} \ll M$, this
approximation implies three important additional conditions.
\begin{enumerate}

\item{The object's radius $R$ should be greater than its classical radius, but much
smaller than the black hole's radius ($P$ shall denote magnetic
monopole and $a$ the specific angular momentum): $q^2/m < R \ll
M+\sqrt{M^2-P^2-a^2}$.}

\item The magnitude of the electromagnetic energy-momentum tensor in the object's immediate vicinity, $|T_{\rm particle}| \sim q^2/R^4$, should be much less than the maximum background spacetime curvature $\sim 1/M^2$ due to the black hole.  Thus  $R\gg \sqrt{qM}$.

\item{The magnitude of the object's interior  energy-momentum tensor should be
much less than the black hole's maximum curvature in its exterior.
Thus $R\gg m^{1/3} M^{2/3}$.}
\end{enumerate}
With condition~1 we insure that the object to be absorbed has
positive energy density. The additional requirement $R \ll M$ makes
the object ``able to fit into the black hole'' and smaller than the
scale set by the curvature, so that finite size effects can be
neglected.  Conditions~2 and 3 keep the ``backreaction" due to the
distortion of the spacetime by the stress-energy of the object
itself negligible, so that the latter's equation of motion is, to
good approximation, the same as in the fixed background geometry of
the black hole.

\subsection{Electric charge or magnetic monopole into Schwarzschild black hole}
\label{Schwarzschild}

Our first example deals with an electrically charged object dropped
from some distance into a Schwarzschild black hole.  The first issue
to resolve is the criterion that prevents the charge from breaking
down the vacuum in its neighborhood (Schwinger effect) and thus
neutralizing itself.  For the breakdown to be suppressed, the
object's electric field just outside it, $\mathcal{E}$, must be well
below the critical field $m_e{}^2/(e\hbar)$, where $m_e$ is the
electron's mass and $e$ is the elementary charge.  Now by condition
2 of Sec.~\ref{sec:setup},  $\mathcal{E}=q/R^2<1/M$. We must thus,
in addition to the three conditions,  require a minimum black hole
mass: $M>e\hbar/m_e{}^2= \alpha^{1/2}\, m_P{}^3/m_e^2\approx
1.05\cdot 10^{39}$ g (the numerical value assumes the real values of
the fundamental constants).  The Hawking radiation of such massive
black hole is indeed negligible.  And the requirement $m\ll M$ still
permits the dropped object to contain many elementary particles and
to be thoroughly classical.

The change in
area during the entire process is given by \Eq{dA} with $m$ replaced by the conserved Killing energy $E$ of the object,
\beq  \delta A = 4\pi(M+E+\sqrt{(M+E)^2-g^2})^2-16\pi M^2.  \eeq
During the assumed adiabatic lowering process, $A$ shall be unchanged.  In addition  we assume the relation $A_i =16\pi M^2$ to remain valid as the particle is being lowered. This just means that $M$, the parameter governing the \textit{background} metric, is taken as constant.  Of course the active gravitational mass of the system changes during the lowering; this corresponds to a metric perturbation of relative order $O(m/M)$ which is then reflected in a commensurate correction to the object's conserved energy. But this last only leads to higher order  corrections to $\delta A$. Therefore, we assume
\beq E = -p_\mu\, \xi^\mu, \eeq
with canonical momentum $p_\mu$ derived from the effective
Lagrangian of the object moving on the fixed curved background, and
$\xi^\mu = (\partial/\partial t)^\mu$ is the timelike Killing
vector.   In our case
\beq E = - m u_t- \half q A^{\rm self}_t. \label{canmom}\eeq
The first term is the fourth component of the  4-momentum of
the particle; the second is the contribution of the object's
electromagnetic vector potential to its own
global energy~\cite{Bekenstein:1999cf,selfforce}.

The self-interaction of a charged object reflects the fact that
Huygens' principle is not satisfied in curved spacetime.  This means
radiative fields propagate not only on characteristic surfaces, but
also inside the light cone as backscattered ``tails". This effect
produces a non-local force term that depends on the metric and past
history of the object. Even when the last is static, the background
curvature distorts its electromagnetic field, which in turn
``backreacts" on it. An alternative intuitive picture of the
self-interaction is that it is due to a surface charge density
induced on the black hole horizon which acts like a conductor.
However, this last intuition does not capture the situation
entirely: the interaction turns out to be \textit{repulsive} and
not, as the analogy would suggest, attractive.

Since the object is initially nearly stationary at, say, $r=r_0$,
its 4-velocity (which is normalized to unity) is $u^a =
\xi^a/|\xi|$, where $|\xi|\equiv \sqrt{-\xi_\mu\,\xi^\mu|_{r=r_0}}$.
$A^{\rm self}_t$ was calculated in
Refs.~\onlinecite{Bekenstein:1999cf,selfforce} and others (see
Appendix A for details) and found to be
\beq A^{\rm self}_t = -\frac{Mq}{r_0^2}, \eeq
so that%
\beq E = m \sqrt{1-2M/r_0}+ \half M q^2/r_0^2.
\label{newE} \eeq

Now according to criterion \reff{Daviesschw} with $m\to E$, the
exact condition for a violation of the area theorem here would be
\beq q^2>4 m M  \sqrt{1-2M/r_0}+ 2 M^2 q^2/r_0^2. \label{crit2} \eeq
To make this more transparent, let us express the dropping
coordinate radius $r_0$ in terms of the proper distance $\ell_0$
from the horizon,
\beq \ell_0 = \int^{2M+\epsilon}_{2M} (1-2M/r)^{-1/2} dr. \eeq
If we assume $\epsilon = r_0 - 2M \ll 2M$, so that we are close to the
horizon in coordinate sense, we find $ \sqrt{1-2M/r_0} \approx \ell_0/4M$ so that
criterion~\reff{crit2} for the area theorem violation becomes
\beq q^2/m \gtrsim 2 \ell_0 \label{ineqSchw}.\eeq

Now the object can be lowered down to the hole only until its center
of mass lies a proper distance $\ell_0 \approx R$ from the horizon,
where it begins to be absorbed by the hole hole. Therefore, since we
require $\ell_0 \gtrsim R$ for the lowering and dropping process,
criterion~\reff{ineqSchw} is inconsistent with condition~1 of
Sec.~\ref{sec:setup}. Were we to neglect the self-interaction
contribution to the energy, the resulting criterion, $q^2/m \gtrsim
\ell_0$, would still be inconsistent with condition~1.  Hence, no
violation of the area theorem can occur when a classical
electrically charged sphere is lowered towards and then dropped into
a Schwarzschild black hole.  By duality, this conclusion is
unchanged if we lower and then drop a magnetically charged sphere.

\subsection{Electric charge into magnetic Reissner-Nordstrom black hole}
\label{sec:monopole}

Now we calculate the horizon area change when an electrically
charged spherical object is lowered towards and then dropped into a
\textit{near-extremal} RN black hole with mass $M$ and magnetic
charge $P$~\cite{duality}. Our motivation for considering this case
is that the spatial geometry becomes throat-like for near-extremal
black holes, allowing the object to have smaller Killing energies at given proper distance from the horizon. And in contrast to the case of an electric charge assimilated by an electric RN,  here there is no direct repulsive interaction between the electric charge and magnetic monopole to contribute to the conserved energy of the
object.

We must still require
$M>\alpha^{1/2} m_P{}^3/m_e^2\approx 1.05\cdot 10^{39}$ g so that the
charge $q$ shall cause no vacuum breakdown.  Since the black hole is
magnetically charged, it cannot cause vacuum breakdown, but it may
polarize the neighboring vacuum with its near magnetic field
$P/M^2$. The polarized vacuum would affect the electrodynamics and
further complicate the already complex situation we analyze. Such
polarization will be suppressed when $P/M^2<m_e{}^2/(e\hbar)$. Since
we  are assuming $P\approx M$, we see that the lower bound we
already required for $M$ automatically suppresses vacuum polarization near the hole.

The exterior geometry is described by the line element and vector
potential
\bea ds^2 &=& -f(r) dt^2 + f(r)^{-1} dr^2 + r^2(\sin^2 \theta\,
d\theta+d\phi^2),\\
\quad A_t&=&P/r,\eea
where $f(r) = 1-\frac{2M}{r}+\frac{P^2}{r^2}$. Now, as before, we
imagine that by some mechanism the object with electric charge $q$
and mass $m$ is slowly lowered from spatial infinity to a fixed
radius $r_0$ outside the black hole, but near the horizon radius,
$r_{+} = M + \sqrt{M^2-P^2}$.

Without loss of generality we assume the object is lowered down the
polar axis ($z$-axis) to $r_0$. Since the configuration contains a magnetic
monopole (in the magnetically charged black hole) and an electric
charge, the electromagnetic and gravitational (in a curved
spacetime) fields together possess a conserved angular momentum $L_z
= qP$ along the $z$-axis, which independent of the distance between object and black hole~\cite{Garfinkle:1990zx,Bunster:2007sn}.
When the object is at spatial infinity, all this angular momentum is
contained in the electromagnetic field, but during the lowering
process, as the electromagnetic field lines penetrate the horizon, the angular momentum is transferred to the gravitational field,  and the hole begins to rotate~\cite{Garfinkle:1990zx,Bunster:2007sn}. We again assume the lowering process is adiabatic, so that during this process the
change in the horizon area will be negligible~\cite{lowering}. Once
the particle reaches $r_0$, it is allowed to fall freely (radially)
into the black hole.

The change in the area is approximately
\bea \delta A &\approx& 4 \pi
\left(\Big(\frac{qP}{M+E}\Big)^2+(M+E)+\sqrt{(M+E)^2-P^2-q^2-\Big(\frac{qP}{M+E}\Big)^2}\right)
\nonumber\\&&- 4 \pi (M+\sqrt{M^2-P^2})^2, \label{dAmonopole}\eea
where we have assumed the final state is a rotating, charged
(Kerr-Newman) solution with angular momentum $qP$ and mass $M+E$.
The energy $E$ contributed by the object is
\beq E \approx m \sqrt{1-2M/r_0+P^2/r_0^2}+
\frac{Mq^2}{2r_0^2}+q \hat{A}_t, \label{Emonopole}\eeq
where we ignore the $O(L_z^2)$ contributions of the hole's rotation to the redshift factor and self-energy.

The self-energy correction in a RN spacetime has the same form as in
Schwarzschild spacetime~\cite{ZelFrolov,Lohiya:1982fp}. The new term
$\hat{A}_t$ originates from the repulsion of the charge $q$ by the
electric dipole induced by the rotation of the magnetically charged
black hole.   Since the charge is brought  near the horizon, we may
assume that the angular momentum parameter of the black hole is  $a
\approx qP/M$. For a magnetic Kerr-Newman black hole, the background
$A_t$ evaluated on the symmetry axis $\theta=0$ is~\cite{Hiscock}
\beq q \hat{A}_t = \frac{qPa}{r_0^2+a^2}= \frac{q^2
P^2}{M(r_0^2+a^2)} \approx \frac{q^2 P^2}{M r_0^2}.\eeq
The proper distance to the horizon is given by integral
\beq \ell_0 = \int^{r_0}_{r_{+}} (1-2M/r+P^2/r^2)^{-1/2}\, dr,
\label{properRN}\eeq
which has the analytic form
\beq \ell_0 =
\sqrt{r_0^2-2Mr_0+P^2}+M\cosh^{-1}\left(\frac{r_0-M}{\sqrt{M^2-P^2}}\right).
\label{properRN1} \eeq
If $r_0-r_{+} \ll r_{+}$ we can expand in a Taylor series as in
Section \ref{Schwarzschild}. However, as the hole becomes more and
more extremal ($\sqrt{M^2-P^2} \rightarrow 0$), the case of most
interest here, this approximation for the proper distance begins to
break down.

Consequently we calculated $\delta A$ in \Eq{dAmonopole} numerically
for a wide range of values of $P/M$, $m/M$, $q/M$ and $r_0/M$.
\textit{In units with} $M=1$ the range studied was $0.9 \leq P\leq
0.99999$, $10^{-12}\leq m\leq 10^{-3}$,  $10^{-4}\surd m\leq q\leq
\surd m$, and $1.00001\leq r_0/r_+\leq 1.05$ .   The upper end of
the range of $q$ is dictated by the condition that the classical
radius of the object, $q^2/m$, be smaller than the black hole radius
(approximately unity).   For the smaller $P$ this condition is
enough to insure that the energy $E$ monotonically increases with
$r_0$ for $r_0>r_+$, which means that the object is always attracted
by the black hole, and  can indeed  be lowered all the way to the
horizon.   As $P$ grows, the $E(r_0)$ curve develops a minimum near
the horizon (with consequent repulsion inside this radius), but even
as $P\rightarrow 1$, this does not happen for $q^2/m<1/3$. Since the
object cannot be entirely made up of electromagnetic energy (as it
would then be unstable), its radius should well exceed $q^2/m$.  Thus for
an object small enough to be dropped into the black hole, the
possibility of its lowering being arrested by the said effect can be
discounted.

The $m$ range was sampled at ten values, each a power of ten greater
than the previous one; the ranges of the other variables were
sampled 100 times each, with jumps of progressively growing (or
decreasing) size in order to maximize coverage of interesting
regions.  For example, the region near $P=1$ was more finely sampled
than that near $P=0.9$, and that near $r_0=r_+$ was covered more
finely than that near $r_0=1.1 r_+$.  For each point of the above
grid we  used \Eq{properRN1} to calculate the proper distance from
$r=r_0$ to the horizon.  The overall range was $0.01165\lesssim
l_0\lesssim 3.2499$. From the above grid of values we discarded
points in conflict with conditions~1-3 of Sec.~\ref{sec:setup}.
Specifically, since we never specified $R$, the particle's radius,
we just demanded that $\ell_0$ exceed each of the quantities $q^2/m,
\surd q$ and $m^{1/3}$. For all parameter combinations meeting these
physical requirements for dropping a finite sized charge into the
black hole without disturbing it strongly, we found that $\delta A >
0$ and so the area theorem is respected.

We stress that for this to be true, the electric dipole-charge
interaction in \Eq{Emonopole} \textit{must} be included.  For example, with
it left out  for $P=0.99999,\, m=10^{-7}$ and $q=10^{-4}$, we find
that $\delta A<0$ for $r_0/r_+<1.058$. At this critical $r_0$,
$\ell_0\approx 7.919$ whereas $\{q^2/m, \surd q, m^{1/3}\}=\{0.1,
0.01, 0.0046\}$.  Thus with the object radius obeying $0.1 \ll R \ll
7.9$, conditions~1-3 are satisfied and we have a violation of the
area theorem. (When only the self-force energy is included,  the
$E(r_0)$ curve has no minimum whatsoever for $q^2/m<1$).

We do not find evidence that the self-interaction term is necessary
for satisfaction of the area theorem (this was also the case for the
Schwarzschild black hole).  Actually, when that term is neglected,
and for $P$ fairly close to $M$, there are formal violations of the
area theorem for $r_0$ in the region where $E(r_0)$ rises with $r_0$, but these
always occur for $l(r_0)$ just marginally larger than $q^2/m$ (7\%
larger in the extreme case).   In view of our above remarks about
the relation between $R$ and $q^2/m$, these cases cannot be counted
as physical violations.  We conclude that for RN black holes the
area theorem is protected by the electric dipole-charge interaction.

At the risk of repetitiousness we mention that analogous conclusions
apply to the case where a \textit{magnetically} charged sphere is
lowered and then dropped into an \textit{electrically} charged RN
black hole.  In that case the area theorem is protected by the
magnetic dipole-monopole interaction.  Breakdown of the near vacuum
by the black hole's Coulomb field and polarization of the vacuum by
the sphere may both be avoided with the same lower bound on $M$
stipulated early in this section.

\subsection{Electric charge into Kerr black hole}
\label{Kerr}

Our third example concerns an object bearing charge $q$ which is
lowered along the symmetry axis of a neutral, nearly extremal Kerr
black hole down to a point where its Killing energy is $E$, and then
dropped in.  The extremal throat-like geometry permits one to
challenge the area theorem to the utmost. Let $M$ be the mass and $a
= J/M$ the angular momentum per unit mass of the initial black hole.
Again, problems stemming from vacuum breakdown by the
object's Coulomb field may be obviated by requiring
$M>\alpha^{1/2} m_P{}^3/m_e^2\approx 1.05\cdot 10^{39}$ g.

When necessary we refer to the Boyer-Lindquist radial coordinate
$r$.  In the envisaged process the angular momentum of the hole is
unchanged. Thus the initial parameter $a$ is transformed into
$a(1+E/M)^{-1}$ so that
\beq \delta A = 4\pi\left(r_f^2-r_+^2+a^2(1+E/M)^{-2}-a^2\right). \label{dAKerr}\eeq
Here $r_f = M+E+\sqrt{(M+E)^2-q^2-a^2(1+E/M)^{-2}}$ and $r_+ \equiv
M + \sqrt{M^2-a^2}$ is the initial Kerr horizon radius. The
conserved Killing energy for a particle on the symmetry axis is
\beq E = m \sqrt{\frac{r_0^2-2Mr_0+a^2}{r_0^2+a^2}}- \half q A^{\rm
self}_t. \label{EKerr}\eeq
In Appendix A we calculate anew the self-interaction term
for an object on the polar axis following the method of
\cite{Bekenstein:1999cf} and find
\beq  -\half q A^{\rm self}_t = \frac{Mq^2}{2(r^2+a^2)}\,. \label{selfKerr1}\eeq
Then we compare this with earlier results of L\'{e}aut\'{e} and
Linet~\cite{Leaute:1982sm}, of Lohiya~\cite{Lohiya:1982fp} and of
Piazzese and Rizzi (PR)~\cite{PiazzRizz}.

Just as in Section \ref{sec:monopole} we here evaluate
formula~(\ref{dAKerr})  numerically using specific values of $m$, $q$, $a$
(in units of $M=1$), and numerically integrating
\beq \ell_0 = \int^{r_0}_{r_+} \sqrt{\frac{r^2+a^2}{r^2-2Mr+a^2}}\, dr
\eeq
(no analytic form is known) to find the proper distance from
the horizon.

The range studied was $0.9 \leq a\leq 0.99999$, $10^{-12}\leq m\leq
10^{-3}$, $10^{-4}\surd m \leq q\leq \surd m$, and $1.000005\leq
r_0/r_+\leq 1.05$.    The upper end of the range of $q$ was chosen
by the consideration set forth in Sec.~\ref{sec:monopole}.   We find
numerically that for $q^2/m\leq 1$ the energy $E(r_0)$ monotonically
increases with $r$ for $r>r_+$, so that  the charge is always
attracted to the black hole, and can be lowered all the way to the
horizon.

The $m$ range was sampled at ten values, each a power of ten greater
than the previous one; the ranges of the other variables were
sampled 100 times each, with jumps of progressively growing (or
decreasing) size so that the region near $a=1$ was more finely
sampled than that near $a=0.9$, and that near $r_0=r_+$ was covered
more finely than that near $r_0=1.05 r_+$.    From the above grid
we again discarded points for which $\ell_0$ did not exceed each of
the quantities $q^2/m, \surd q$ and $m^{1/3}$, thus securing
compliance with conditions~1-3.  We uncovered no violations of the
theorem for the residual points.

When the self-force energy, \Eq{selfKerr1}, is not included,
violations of the area theorem are found. For example, with $m= 3
\times 10^{-8}$, $q = 2 \times 10^{-5}$ and $a=0.9999$, $\delta A <
0$ when $r_0 < 1.0145 $, that is, when the object is dropped a
proper length $\ell_0 < 0.3199$ from the horizon. Since $\{q^2/m,
\surd q, m^{1/3}\}=\{0.0133, 0.00447, 0.00311\}$,  if the size of
the object obeys $ 0.0134 <  R \ll 0.31$, conditions~1-3 are
met, albeit marginally.  The self-force  is obviously an
essential part of the workings of the area theorem.

We also examined the situation when the object's energy was
calculated using the alternative form for the self-force (\Eq{alter}
below) worked out  by PR~\cite{PiazzRizz}, to which corresponds the
energy correction \beq E_{\rm PR} =
\frac{q^2(M+r_0)}{2(r_0^2+a^2)}+\half \frac{q^2}{a} {\rm
Arctan}(r/a) - \frac{\pi q^2}{4a}\,, \label{selfPR} \eeq instead of
that in \Eq{selfKerr1}. We easily found violations of the area
theorem.  For the same parameters $a, e$ and $m$ as above, we find
that $\delta A < 0$ when $r_0 < 1.0144 $, that is, when the object
is dropped a proper length $\ell_0 < 0.2718$ from the horizon.
Obviously conditions~1-3 can be satisfied in this case for $0.0134 <
R \ll 0.271$.  These and similar violations of the area theorem
constitute independent evidence that the PR formula for the
self-force is incorrect.

\section{Lowering an elementary Charge}
\label{ep}

In this section we suppose that the above \textit{gedanken}
experiments of Sec.~\ref{sec:lowering} are carried out with an elementary particle of mass $m$,
and charge $e$. To be concrete we work in the Schwarzschild case,
where we can make use of the analytic results in Section
\ref{Schwarzschild}. 

The criterion~\reff{ineqSchw} for $\delta A < 0$  translates into
\begin{equation}
e^2/M=\alpha\cdot \hbar/m\gtrsim 2\ell_0.
\end{equation}
As discussed earlier, in a world where $\alpha>1$ the spatial extent of the particle might be given by its classical radius $e^2/m$.  In such eventuality the last criterion could not be satisfied since it requires the still suspended particle's center to be closer to the horizon than its own radius.  The conclusion might be that no violation of the area theorem occurs,  but clearly one cannot draw any bound on $\alpha$.

If instead the particle's effective size is about a Compton wavelength, its center cannot be deposited closer to the horizon than that distance, so
$\ell_0\gtrsim\hbar/m$. Thus violation of the area law would entail
$\alpha\gtrsim 2$. Assuming Hawking radiation effects can be
neglected, we may conclude that the area theorem implies the bound %
\begin{equation}
\alpha\lesssim 2. \label{ineq}
\end{equation}
(In the RN and Kerr cases this bound should also be of order
unity when we include the dipole repulsion and self-forces).
Apart from the factor 2, this is bound (\ref{Daviesh2}) inferred by
Davies from the freely falling charge experiment. The weaker constraint here
results from our inclusion of the self-force energy. Note
that Davies would also have obtained $\alpha\lesssim 2$ had he
demanded that the black hole \textit{radius} exceeds the Compton
length.

Now as we stressed in Section \ref{sec:crit}, in a world where
$\alpha \gtrsim 1$, the area theorem is unreliable because
the charge will  polarize the vacuum, so that the expectation
value of the stress-energy tensor operator need not then satisfy the weak energy condition. In an attempt to settle the issue here, we again appeal to the more reliable GSL and use  Eq.~(\ref{Hawkcont})
for the net contribution of the Hawking radiation to the change in
total entropy (black hole plus exterior).  We repeat the derivation of \Eq{GSL2}, but this time with $m$ replaced by the $E$ of \Eq{newE} with $ \sqrt{1-2M/r_0} \approx \ell_0/4M$ and $r_0\approx 2M$.  We find that the GSL
requires
\beq \frac{\gamma \Delta t}{1960 M}+\frac{\pi}{\hbar}(2\ell_0 m-
e^2) \geq 0.
\label{GSL3} \eeq

What to take for the disappearance time $\Delta t$?  We are allowed to lower down the charge to within roughly a Compton wavelength from the horizon, so $\hbar/m$ is a lower bound on $\Delta t$.   And because $M$ sets the scale of the gravitational field, a geneous upper bound should be a few times $M$.  We conclude that  the first term in \Eq{GSL3} is small compared to unity, and thus negligible compared to $2\pi\ell_0 m/\hbar$.  Hence the GSL gives $e^2/\hbar\lesssim  2\ \ell_0/\lambda_{\rm C}$; since we can arrange for  $\ell_0$ to approach $ \lambda_{\rm C}$, we recover  inequality~(\ref{ineq}).

Are there any relevant effects which might change this conclusion?  
One possibly important effect is Unruh-Wald or
quantum buoyancy~\cite{Unruh:1982ic} . This buoyancy comes about
because a suspended, hence accelerated, object perceives the quantum vacuum outside a
black hole as an ``atmosphere" of thermal radiation. Since the local
temperature at the bottom of the object is higher than that at its
top (due to the varying redshift factor), the associated pressure gradient exerts an outwardly directed force on the object.   Consequently  the conserved energy  $E(r_0)$ receives an extra positive contribution from the work done by the buoyancy on the object. In the classical object examples in Sec.~\ref{sec:lowering} this contribution is negligible, but in any case makes it more difficult to violate the area theorem.

For an elementary point charge in the thermal atmosphere the fluid description of radiation which underlies Unruh and Wald's
original calculation of buoyancy~\cite{Unruh:1982ic} must be replaced by one based on the momentum transfer to the charge due to Compton scattering of atmosphere photons.  Thus, it is not obvious whether the neglect of this effect
is also similarly justified for the point particle case.
In Ref.~\onlinecite{Bekenstein:1999bh} one of us calculated, in Schwarzschild spacetime, the average
repulsive force measured at infinity on a suspended  elementary charge due to Compton scattering finding
\begin{equation}
{\bf f}= \frac{\hbar\, \hat{\bf z}}{270 \pi M}~\frac{\ell_0{}^2\,
r_c{}^3}{\left(\ell_0{}^2-r_c{}^2\right)^3}\label{elemforce}
\end{equation}
where ${\hat{\bf z}}$ is a unit vector pointing away from the
horizon in an orthonormal frame centered at the particle. For
$\alpha \ll 1$, $\ell_0>\lambda_{\rm C} \gg r_c$ and one can rewrite
(\ref{elemforce}) as
\begin{equation}
{\bf f}\approx (2/135\pi)\alpha^3 (\lambda_{\rm C}/\ell_0)^4\, {\bf
f}_{\rm grav}, \label{approxelemforce}
\end{equation}
where ${\bf f}_{\rm grav}=\hat{\bf z}\,m\,(4M)^{-1}$ is the
gravitational force as measured at infinity. Thus for the observed
$\alpha \approx 1/137$, the repulsive force is clearly
negligible compared to the gravitational attraction since $\ell_0 >
\lambda_{\rm C}$ during the lowering.

However, for our case of interest, $\alpha \gtrsim 1$,
Eq.~(\ref{elemforce}) is unreliable because it is based on a leading
order (tree-level in $\alpha$) computation of the differential
scattering cross-section, a procedure not justified when QED is strongly
coupled.  It would be useful to determine the momentum transfer due to Compton scattering to higher order, but this calculation is beyond the scope of this paper.  Instead, let us try gain some insight by naively
extrapolating Eqs.~(\ref{elemforce}) and (\ref{approxelemforce}) into
the strongly coupled regime $\alpha \sim 1$. \Eq{approxelemforce} suggests the repulsion is \textit{not} strongly
suppressed compared to the gravitational attraction when
$\ell_0 \sim  \lambda_{\rm C}$, while Eq.~(\ref{elemforce}) suggests that the repulsive force actually diverges as the dropping distance $\ell_0$
approaches a Compton length from the horizon.

Though not definitive, this argument suggests that the
interaction of the elementary particle with the thermal atmosphere
makes a significant contribution to the energetics of the lowering
process when $\alpha \gtrsim 1$. Due to this contribution the charge will probably  reach a floating point near the horizon~\cite{Unruh:1982ic,Bekenstein:1999bh}. At any rate, the change in energetics will weaken the implied constraint on the fine structure constant to an extent we cannot determine with the facts available.  It is even possible that by preventing
GSL violation, the buoyancy may, in fact, remove the constraint completely.

\section{Summary and discussion}
\label{discussion}

We reassessed the \textit{gedanken} experiment used by
Davies~\cite{Davies:2007wa} to set constraints on the value of the
fine structure constant $\alpha=e^2/\hbar$  and on the spatial extent of magnetic monopoles. We showed that in order for constraints to
follow from the area theorem, one must assume the object dropped from
infinity must be a quantum ``point" particle whose localization
scale is its Compton wavelength $\lambda_{\rm C} = \hbar/m$. If instead an extended classical charged object is dropped, violation of the area theorem transpires only when the object violates the weak energy condition. Hence the only thing we learn in this case is the obvious requirement that realistic classical objects must obey the classical energy conditions.
However, for an intrinsically quantum elementary charge,
the violation of Davies' constraint $e^2 < \hbar$ implies that the charge strongly polarizes the surrounding vacuum, raising doubts as to whether the weak energy condition is maintained.

Therefore we are forced to fall back on the full GSL.  To analyze the
effect of the Hawking contribution to the entropy change, $\delta S_{\rm ext}$, we imagined the \textit{gedanken} experiment to be performed by dropping a series of particles into the hole at a certain rate. If the (strong) assumption is made that no instabilities arise when the dropping rate is tuned to the maximum value required by the gedanken experiment, Davies' bound on $\alpha$ is only moderately weakened.

With the Davies experiment in mind we then examined a new class of
\textit{gedanken} experiments in which classical charged test objects are lowered towards and then dropped into black holes of Schwarzschild, RN or Kerr type. It turns out that this apparently simpler type of experiment does tell us something interesting about black hole
physics. When an electrically (magnetically) charged object is
dropped into a near-extremal magnetic (electric) RN black hole, the subtle correction to the object's Killing energy due to the electric (magnetic) dipole repulsion turns out to be crucial for consistency between the area theorem and the weak energy condition. We show numerically that when this effect is included, the black hole area decreases only in cases where the object is more compact than its classical radius, under which circumstance the area theorem is expected to fail anyway. Similarly, when a charged object is lowered along the symmetry axis of a Kerr black hole and then dropped in, it is the electric self-force on the particle that prevents violation of the area theorem.

Thus the classical area theorem is only valid for
near-extremal black holes when ``backreaction" effects like
self-forces and the dipole repulsion in the RN case are properly
taken into account. A naive analysis that only considers the
gravitational and Coulomb forces can disclose apparent violations. In
this respect it is interesting to note that similar backreaction
effects also seem to play an important role in maintaining the
cosmic censorship hypothesis when one attempts to destroy the
horizon by overcharging or over-spinning a near-extremal black hole.
In the test particle limit cosmic censorship can apparently be
violated in some cases by overcharging~\cite{Hubeny:1998ga}, but it
appears that higher order backreaction effects prevent the horizon's
destruction, and the concomitant violation of the GSL~\cite{Hod}.

In Section \ref{ep} we considered \textit{gedanken} experiments
where an elementary particle is lowered to nearly a Compton
wavelength away from the horizon and then dropped in. We
obtain what seems to be a very robust bound on $\alpha$. On the
other hand, we showed that a neglected quantum effect, the
Unruh-Wald quantum buoyancy felt by the charge, as it is held at
rest in the ``thermal atmosphere" near the horizon, is likely to
make a significant contribution to its conserved energy. This effect could protect the GSL at large $\alpha$, and make
the bound unnecessary. Again, it is interesting to note that the
Unruh-Wald buoyancy has also been shown to be key in upholding
cosmic censorship and preventing the destruction of the black hole
horizon in a similar experiment where an electric charge is lowered
to an electric RN hole~\cite{Hod:2000bi}.

In light of the above, let us reconsider the bound on $\alpha$ that follows from Davies' original experiment. There
the particle is dropped from rest at infinity and would seem to freely
fall into the hole, so that the thermal atmosphere should have no effect on it.  Note however that the radiation reaction on the charge due to
self-field effects was neglected by Davies (who correctly notes that any radiated energy that escapes entering the hole will only make the bound on $ \alpha$ the stronger). This neglect is actually justified only when the classical radius of the particle is much smaller than the radius of curvature of the background spacetime: $e^2/m \ll M$ \cite{Hubeny:1998ga}.  But the Davies argument requires  $e^2/m =\alpha\cdot \hbar/m \sim 4\alpha M$.  Thus if we want to allow \textit{a priori} the possibility that $\alpha>1$, radiation reaction may be be important.  One effect of it would be to make the charge move in a radial non-geodesic trajectory.  Because the charge would then be accelerating, some variant of quantum buoyancy might act on it (although not exactly Unruh-Wald buoyancy from thermal radiation).  Near the horizon  the new buoyancy may be strong, so it may actually stop the charge and make it rebound outward.  It is beyond the scope of this paper to clarify this issue by calculations.

Thus in both methods for dropping the elementary charge, subtle effects---quantum buoyancy in one case  and radiation reaction together with quantum buoyancy in the second---``muddy the waters'', and it is not at all clear whether any
bound on the fine structure constant is required by the GSL. In view of
the lessons of Section \ref{sec:lowering} in the classical regime, it is not unreasonable to speculate that when all effects are properly considered, the full GSL will be found to be respected without the need for a fundamental bound on $\alpha$.  As in its early history, the black hole area  theorem may help us to understand new physical effects.

\section*{Acknowledgements} This work was supported by grant 694/04 of the Israel Science Foundation and by the Lady Davis Foundation at Hebrew University. We thank S. Green for
remarks on the preprint that helped us correct an error in the first
version of the results of Sec.~\ref{Kerr}.

\section*{Appendix A: Calculation of the self-energy in Kerr geometry}
\label{AppendixA}
\renewcommand{\theequation}{A.\arabic{equation}}
\setcounter{equation}{0}

Here we extend the method of~\cite{Bekenstein:1999cf} to calculate
the self-interaction correction,
\beq E_{\rm corr} = -\half q A^{\rm self}_t, \eeq
to the conserved energy of a stationary test object on the symmetry
axis of the Kerr geometry. The procedure starts with Maxwell's
equations for the vector potential $A_\mu$ with a point charge
source at $r=r_0$ on the axis. Fortunately, an analytic solution was
found by L\'{e}aut\'{e} \cite{Leaute1977}, who was able to extend
the earlier solution of Copson and Linet~\cite{Copson,Linet:1976sq}
in the Schwarzschild background to the Kerr case. In Boyer-Lindquist
coordinates $(t,r,\theta,\phi)$,
\bea A_t(r,\theta) &=& -\frac{q}{\Sigma_0 \Sigma}\left[(r_0 r+a^2\cos
\theta) \left(M+\frac{(r-M)(r_0-M)-(M^2-a^2)\cos \theta}{R}\right) \right.\nonumber\\
&& \left. + a^2(r-r_0 \cos \theta) \frac{(r-M)-(r_0-M) \cos
\theta}{R}\right], \label{AKerr}\eea
where $\Sigma_0 = r_0^2+a^2$, $\Sigma = r^2+a^2$, and $R =
(r-M)^2+(r_0-M)^2-2(r-M)(r_0-M)\cos \theta-(M^2-a^2) \sin^2 \theta$.

We first re-express this solution in isotropic coordinates. In
Schwarzschild geometry one can define a new radial coordinate
$\varrho$ which makes the spatial 3-metric conformally flat and
isotropic. While no such possibility exists for the Kerr spacetime,
the coordinate change $r =
\bar\varrho+M+\frac{(M^2-a^2)}{4\bar\varrho}$ does put the metric
into the form
\beq ds^2 = -A(\bar\varrho,\theta)dt^2 + B(\bar\varrho,\theta)[d\bar\varrho^2+\bar\varrho^2 d\theta^2] -
C(\bar\varrho,\theta) dt d\phi + D(\bar\varrho,\theta) d\phi^2, \eeq
which is isotropic in the $(\bar\varrho,\theta)$ plane. Since the
solution in question has only $A_t, A_\phi$ components we may
transform $A_t$ to the new coordinates by just re-expressing  the
function $A_t(r,\theta)$ as a function of $\bar\varrho$ and $\theta$
which we shall denote $A_t(\bar\varrho,\theta)$, and then use the
substitutions
\bea \bar\varrho \cos \theta &\rightarrow&   \bar\varrho_0+\varrho \cos \vartheta\\
\bar\varrho &\rightarrow& \sqrt{\bar\varrho_0^2+\varrho^2+2\bar\varrho_0 \varrho \cos \vartheta}
\eea
to transform to a new set of isotropic coordinates
$(\varrho,\vartheta,\phi)$ centered on the location of the object's
center of mass $\bar\varrho_0$. Expanding $A_t(\varrho,\vartheta)$
in powers of $\varrho$, we find
\beq A_t = -q |\xi| \epsilon^{-1}- \mathcal A +
\mathcal B \cos\vartheta + O(\varrho) + \cdots, \label{Aseries}\eeq
where $|\xi| = \sqrt{\frac{r_0^2-2Mr_0+a^2}{r_0^2+a^2}}$, $\epsilon
= |\xi|^{-1} \varrho$ is proper distance from the center of mass,
and $\mathcal A$ and $\mathcal B$ are constants depending on $M$,
$a$, $q$, and $\varrho_0$. The first term in (\ref{Aseries}) is the
Coulomb potential of a point charge ($t$ component) redshifted to
the location $r_0$ in the gravitational field of the hole. Assuming
the object is spherical, in the limit $\varrho \rightarrow 0$ this
last term renormalizes the rest mass of the object
\beq m_{\rm rem} = m + \lim_{\epsilon \rightarrow 0} q^2 |\xi|/2\epsilon,
\eeq
while first and higher order contributions in $\varrho$ vanish. The
third term $\mathcal B \cos\vartheta$ vanishes when averaged over
the angular direction $\vartheta$ in our isotropic space. Converting
the complicated expression for $\mathcal A$ in terms of $\bar
\varrho_0$ back to Boyer-Lindquist $r_0$, one is left with just
\beq E_{\rm corr} = \frac{Mq^2}{2(r_0^2+a^2)}. \label{selfKerr}\eeq

This result is consistent with the literature on the subject.
L\'{e}aut\'{e} and Linet~\cite{Leaute:1982sm} and Lohiya
\cite{Lohiya:1982fp} found a repulsive force along the polar axis
\beq \overline{F} = \frac{M q^2 r_0}{(r_0^2+a^2)^2}. \eeq
Their method was to re-express the vector potential (\ref{AKerr}) in
coordinates where the gravitational field in the vicinity of the
particle is locally homogeneous (a local Rindler frame) to first
order. In these coordinates the potential has the form of that for
an accelerated point charge in flat spacetime plus an additional
term due to the background curvature. Using the formula for the
electric field $E_i = \partial_i A_t$ and averaging over all
directions yields the result above. In contrast, Lohiya calculated
the electric field due to the point charge along the symmetry axis
of Kerr in Boyer-Lindquist coordinates. In the limit $r \rightarrow
r_0$, the result is again a Coulomb piece whose corresponding energy
renormalizes the mass of the particle, the self-force term mentioned
earlier, and terms depending on the sign of $r-r_0$ which average
out to zero if we imagine the charge is assembled in a spherically
symmetric manner.

The contribution that L\'{e}aut\'{e}, Linet and Lohiya's self-force
makes to $E$ can be found as follows. The work done in a local
orthonormal frame at $r=r_0$ when the charge is displaced  a proper
distance $d \ell$ is
\beq d \overline{W} = \overline{F(r_0')}\, d\ell. \eeq
At infinity this work is measured as $d E = |\xi| d
\overline{W}$ because the with redshift factor is $|\xi|$. Thus,
\beq E = \int^{\infty}_{r_0} |\xi| \overline{F} d\ell,
\label{forceintegral} \eeq
where $d\ell = |\xi|^{-1} dr'_0$ along the axis. Since the redshift
factors cancel, this is just an integral of the self-force from $r_0$
to $\infty$, which agrees with our result \Eq{selfKerr}.

Piazzese and Rizzi~\cite{PiazzRizz}
reexamined the Kerr self-force problem and came up with a different
self-force,
\beq \overline{F} = \frac{q^2 (Mr_0-a^2)}{(r_0^2+a^2)^2},
\label{alter}
 \eeq
again by differentiating $A_t$ to find an electric field in
coordinates where the gravitational field is locally homogeneous.
Using (\ref{forceintegral}) one now finds the energy correction to
be given by~\Eq{selfPR} instead of \Eq{selfKerr}. As we remarked in
Section \ref{Kerr}, this alternative energy correction is
inconsistent with the area theorem.

Independently of the above, we believe PR's analysis is in error.
They define the proper displacement $d \ell$ in the formula
\beq \overline{F} = \frac{1}{|\xi|} \frac{dE}{d\ell} \eeq
in terms of the Boyer-Lindquist coordinate difference $r-r_0$. In
this formulation an extra term proportional to $|r-r_0|$ in the
expansion for $A_t$ appears to contribute to the force. However, the
correct physical picture of the force involves a displacement of the
charge location  $r_0$  itself. In other words, one should take the
limit as $r \rightarrow r_0$ in $A_t$ first, and only then
differentiate with respect to $r_0$ to find the self-force.

\section*{Appendix B: Must a cable along the Kerr polar axis always snap?}
\label{AppendixB}
\renewcommand{\theequation}{B.\arabic{equation}}
\setcounter{equation}{0}

One possible inconsistency in our {\it gedanken} experiments is an
instability in the lowering and dropping process. Maybe the cable
always snaps before we can place the object at rest at the desired radius
$r_0$ near the horizon? In Section \ref{sec:setup}, we briefly
described the cable model of Ref.~\onlinecite{Fouxon:2008pz}, which
indicates it is possible, at least in principle, to complete a
lowering and dropping process in the Schwarzschild geometry. There
the cable was considered to be conical, made up of thin radial
fibers filling the portion of space defined by some solid angle with
vertex near the black hole. With this cable design the enormous
tension needed to hold a load stationary near the horizon can be
distributed over a steadily increasing cross-sectional area.

However, the situation is more complicated in the Kerr spacetime. We
must also consider the effect of the hole's rotation on the cable
system, the so-called ``dragging of inertial frames".  When the
cable is placed along the symmetry axis ($\theta=0$), the rigid
fiber elements will begin to rotate with the black hole in the
$\phi$ direction at an angular velocity that depends on the distance
from the horizon,
\beq \omega(r) = \frac{-g_{t\phi}}{g_{\phi \phi}}=
\frac{2Mra}{(r^2+a^2)^2}, \eeq
(Boyer-Lindquist coordinates). As the fibers twist around each
other, the stress tensor of the cable will come to include not only
the radial tension and mass density, but also shearing, torsional
stresses ($\phi\phi$ component) and a flow of mass-energy in the
$\phi$ direction.

To circumvent this problem we propose a new, less rigid, suspension
in which the fibers are allowed to rotate freely, as dictated by the
geometry, without twisting. The object of mass $m$ is attached to
the lowest of a system of separate light rigid disks which are
orthogonal to the symmetry axis and concentric with it; the
uppermost disk is attached to some distant fixed structure.  Each
disk has a number of fibers affixed to its top surface, and each
such fiber is suspended by a bearing from a circular groove in the
next disk up in such way that the suspended end can slide without
friction.  This arrangement prevents the shearing between fibers; at
each tier the fibers will rotate with respect to infinity with the
local $\omega(r)$.  We suppose the number of fibers grows from tier
to tier, as does the number of grooves per disk, replicating the
conical structure already mentioned. We shall require that the full
width of the suspension lie within a solid angle of small
opening $\theta$ with apex at
the suspended object.  In this way we can proceed to the limit
$\theta\to 0$ at the end of our calculations.

In the Kerr spacetime it is convenient to work with the field of
orthonormal tetrad frames corresponding to locally non-rotating
observers; these will rotate with the fibers at the angular velocity
$\omega(r)$. The basis vectors for this orthonormal frame are, in
terms of Boyer-Lindquist coordinates,~\cite{Bardeen:1972fi}
\bea \mathbf{e}_{\hat{t}} &=& (\frac{\mathcal{A}}{\Sigma
\Delta})^{\half} \frac{\partial}{\partial t}+
\frac{2Mar}{(\mathcal{A} \Sigma \Delta)^{\half}}\frac{\partial}{\partial \phi}\,,  \\
\mathbf{e}_{\hat{r}} &=& (\frac{\Delta}{\Sigma})^{\half} \frac{\partial}{\partial r}\,, \nonumber\\
\mathbf{e}_{\hat{\theta}} &=& (\frac{1}{\Sigma})^{\half} \frac{\partial}{\partial \theta}\,, \nonumber\\
\mathbf{e}_{\hat{\phi}} &=&
(\frac{\Sigma}{\mathcal{A}})^{\half}\frac{1}{\sin \theta}
\frac{\partial}{\partial \phi}\,,\nonumber \eea
where $\Delta=r^2+a^2-2Mr$, $\mathcal{A} = (r^2+a^2)^2-a^2 \Delta
\sin^2 \theta$ and $\Sigma = r^2+a^2 \cos^2 \theta$. In this frame
there is no energy flux in the $\phi$ direction, and the only
non-zero components of the stress tensor, averaged over many fibers,
are the mass density $\rho$ and the tension per unit cross-section
$S$, both of which we naturally take to depend only on $r$ since the
fibers run parallel to the symmetry axis:
\bea T^{\hat{t} \hat{t}} &=&\rho(r), \nonumber\\
T^{\hat{r} \hat{r}} &=& S(r). \eea
From these we may calculate $T^{\mu\nu}$ in the Boyer-Lindquist
coordinate frame.

A stationary configuration of matter outside the hole obeys the
conservation law
\beq T^{\nu}_{\mu;\nu}=\frac{(\sqrt{-g} T_\mu^\nu),_\nu
}{\sqrt{-g}}-\frac{1}{2} g_{\alpha\beta},_\mu T^{\alpha\beta}= 0.
\eeq
Along the symmetry axis only the $\hat{r}$ component of this gives a
nontrivial condition:
\beq \frac{dS}{dr} = \frac{M\rho(r)(a^2-r^2)+S(r)(a^2 M+3r^2 M-2r^2
a-2r^3)}{(r^2-2Mr+a^2)(r^2+a^2)} \label{diffS} \eeq
By assuming that the suspension flares upward so as to fill a small solid angle around the axis, we may continue to use this equation slightly off axis.  In this case $S$ signifies stress along the fibers.
Our goal is to show that there is a (non-singular) solution to this
equation that satisfies the appropriate boundary conditions and the
weak energy condition, $|S|/\rho \leq 1$.

Ref.~\onlinecite{Fouxon:2008pz} assumed a constant
(average) proper density $\rho=\rho_0$; this implies 
infinite sound speed in the suspension apparatus.  We avoid this unphysical assumption. Assuming that the stretching is adiabatic, the constitutive relation should be $S=-\mathcal{S}(\rho)$, where $\mathcal{S}(\rho_0)=0$ for the density $\rho_0$ of the unstressed apparatus,  and $\mathcal{S}(\rho)>0$ for $\rho>\rho_0$.  (Recall that $S<0$ for a
suspension under tension; in this state the density should exceed
$\rho_0$ since elastic energy adds to $\rho$). The squared speed of sound along the apparatus axis is $\mathcal{S}'(\rho)$; we thus assume $\mathcal{S}'(\rho)< 1$ for all $\rho\geq \rho_0$.  It is immediately clear that $\mathcal{S}(\rho)< \rho$.  Thus $|S|< \rho$ so that the weak energy condition is satisfied in as a result of the causality assumption.  

To make calculations tractable we shall specialize to a linear $\mathcal{S}$, that is
\beq S = K\cdot(\rho_0-\rho), \label{eos} \eeq
where $K$ is the squared speed of sound.    Consequently we assume $K <
1$ to preserve causality; again this immediately implies that $|S|/\rho <
1$.  Therefore the weak energy condition will be automatically
satisfied, provided there exists a non-singular solution for the
stress.

Using \Eq{eos} we eliminate $\rho(r)$ in favor of $S(r)$ in
\Eq{diffS}. The general solution to the resulting differential
equation for $S(r)$, with $C$ an integration constant, is
\beq S(r) = (r^2-2Mr+a^2)^{-\gamma}(r^2+a^2)^{\gamma-1}
\left[\int^{r}_{r_0} M\rho_0 (a^2-r'^2) (r'^2+a^2)^{-\gamma}
(r'^2-2Mr'+a^2)^{\gamma-1} dr' + C\right],\label{tension} \eeq
where $r=r_0$ marks the lower end of the suspension and $\gamma =
\frac{K-1}{2K}< 0$ given our causality restriction. The
term involving the integral describes the average stress (tension
per unit area) due to the weight of fibers and disks, $S_{\rm susp}$. The second term is due to the load on the end of the
suspension. Using \Eq{tension}, we can solve for the integration
constant $C$ in terms of $S(r_0)$:
\beq C = S(r_0)(r_0^2-2Mr_0+a^2)^{\gamma}(r_0^2+a^2)^{\gamma-1}. \eeq
For a suspended point mass $m$, $S(r_0) = -m
a^{\hat{r}}/\mathfrak{a}$, where $a^{\hat{r}}$ is the radial
acceleration in the orthonormal frame at $r_0$ and $\mathfrak{a}$ is
the cross-section of the suspension at the bottom. We thus find
\beq S_{\rm load}
=\frac{mM(a^2-r_0^2)(r_0^2-2Mr_0+a^2)^{\gamma-\half}}{\mathfrak{a}(r_0^2+a^2)^{\gamma+\half}
(r^2-2Mr+a^2)^{1-\gamma}(r^2+a^2)^{\gamma}}.
\label{load} \eeq
Note that this stress is negative (as befits a tension) and, for
sufficiently large $r$, its magnitude decreases monotonically as $r^{-2}$ because the suspension's cross-section increases.

Generically $S_{\rm load}$ becomes very large as $r_0$ approaches the horizon radius $r_{+}$. In (\ref{eos}) this corresponds to a cable density $\rho \gg \rho_0$. In this regime the validity of
the linear constitutive relation is doubtful. However, by taking the
suspended mass $m$ sufficiently small on scale $M$, and by employing a suspension with an already large cross-section $\mathfrak{a}$ at its bottom, the large stresses can be controlled. 

Hence, our remaining task to check that there are non-singular solutions for the stress. Inspection of the integral piece of (\ref{tension}) indicates no obvious problems; for example, for large $r$, $S(r) \sim r^{-1}$. We have numerically integrated in the cases where $K < 1 $ and for various values of $\rho_0$ to deduce $S_{\rm susp}(r)$ near the horizon. We scanned through a series of dropping points $r_0>r_{+}$ near the horizon and found well-behaved solutions throughout. Thus it seems the weak energy condition can be satisfied during the lowering process, so there is no reason \textit{of principle} that will force our suspension to break. We conclude there is no inconsistency in the lowering and dropping process envisaged in Sec.~\ref{Kerr}.


\begin{thebibliography}{99}

\bibitem{Hawking:area}S.~W.~Hawking, Phys. Rev. Lett. \textbf{26}, 1344 (1971).

\bibitem{Bek:super}J.~D.~Bekenstein, Phys. Rev. D \textbf{7}, 949 (1973).

\bibitem{Bek:GSL}J.~D.~Bekenstein, Phys. Rev. D \textbf{9}, 3292 (1974).

\bibitem{Davies:2007wa}
  P.~C.~W.~Davies,
  Int.\ J.\ Theor.\ Phys.\  {\bf 47}, 1949 (2008)
  [arXiv:0708.1783 [gr-qc]].

\bibitem{Zeldovich}
Ya.~B.~Zel'dovich and V.~S.~Popov, Sov.\ Phys.\ Usp.\ {\bf 14}, 673
(1972).

\bibitem{Hawking:rad}
S.~W.~Hawking, Comm.\ Math.\ Phys.\ {\bf 25}, 152 (1972).


\bibitem{Fouxon:2008pz}
  I.~Fouxon, G.~Betschart and J.~D.~Bekenstein,
  Phys.\ Rev.\  D {\bf 77}, 024016 (2008)
  [arXiv:0710.1429 [gr-qc]].

\bibitem{membrane}
{\it Black Holes: The Membrane Paradigm}, edited by
K.~S.~Thorne, R.~H.~Price, and D.~A.~MacDonald (Yale University Press, London, 1986).

\bibitem{Bekenstein:1999cf}
  J.~D.~Bekenstein and A.~E.~Mayo,
  Phys.\ Rev.\  D {\bf 61}, 024022 (1999)
  [arXiv:gr-qc/9903002].

\bibitem{selfforce}
Early calculations of the self-force were by A.~Vilenkin,
  Phys.\ Rev.\  D {\bf 20}, 373 (1979) as well as A.~G.~Smith and C.~M.~Will,
  Phys.\ Rev.\  D {\bf 22}, 1276 (1980).

\bibitem{duality}By electromagnetic duality
this case is completely equivalent to a magnetic charge being
dropped into an electric RN black hole.

\bibitem{Garfinkle:1990zx}
  D.~Garfinkle and S.~J.~Rey,
  Phys.\ Lett.\  B {\bf 257}, 158 (1991).

\bibitem{Bunster:2007sn}
  C.~Bunster and M.~Henneaux,
  PNAS, {\bf 104} (30), 12243 (2007).

\bibitem{lowering}An equivalent setup is to think of the particle as having just been
``placed" at rest at $r_0$ outside the black hole. However, in this initial
configuration the hole must have some rotation to be consistent with the adiabatic
lowering picture (and the initial value constraint equations).

\bibitem{ZelFrolov}
A.~I.~Zel`nikov and V.~P.~Frolov, Zh. Eksp. Teor. Fiz. {\bf 82}, 321
(1982) [Sov. Phys. JETP {\bf 55}, 191 (1982)].

\bibitem{Lohiya:1982fp}
  D.~Lohiya,
  J.\ Phys.\ A  {\bf 15}, 1815 (1982).

\bibitem{Hiscock}
W.~A.~Hiscock, Ann. Phys. {\bf 131}, 245 (1981).

\bibitem{Leaute:1982sm}
  B.~L\'{e}aut\`{e} and B.~Linet,
  J.\ Phys.\ A  {\bf 15}, 1821 (1982).

\bibitem{PiazzRizz}
F.~Piazzese and G.~Rizzi, Gen.\ Rel.\ Grav.\ {\bf 23}, 403 (1991).

\bibitem{Unruh:1982ic}
  W.~G.~Unruh and R.~M.~Wald,
  Phys.\ Rev.\  D {\bf 25}, 942 (1982).

\bibitem{Bekenstein:1999bh}
  J.~D.~Bekenstein,
  Phys.\ Rev.\  D {\bf 60}, 124010 (1999)
  [arXiv:gr-qc/9906058].

\bibitem{Hubeny:1998ga}
  V.~E.~Hubeny,
  Phys.\ Rev.\  D {\bf 59}, 064013 (1999)
  [arXiv:gr-qc/9808043].

\bibitem{Hod}
  S.~Hod,
  Phys.\ Rev.\ Lett.\  {\bf 100}, 121101 (2008)
  [arXiv:0805.3873 [gr-qc]];   S.~Hod,
  Phys.\ Rev.\  D {\bf 66}, 024016 (2002)
  [arXiv:gr-qc/0205005].

\bibitem{Hod:2000bi}
  S.~Hod and T.~Piran,
  Gen.\ Rel.\ Grav.\  {\bf 32}, 2333 (2000)
  [arXiv:gr-qc/0011003].


\bibitem{Leaute1977}
B.~L\'{e}aut\`{e}, Ann.\ Inst.\ H.\ Poincar\`{e} {\bf 27}, 167 (1977).

\bibitem{Copson}
E.~Copson, Proc.\ Roy.\ Soc.\ London A {\bf 118}, 184 (1928).

\bibitem{Linet:1976sq}
  B.~Linet,
  J.\ Phys.\ A  {\bf 9}, 1081 (1976).

\bibitem{Bardeen:1972fi}
  J.~M.~Bardeen, W.~H.~Press and S.~A.~Teukolsky,
  Astrophys.\ J.\  {\bf 178}, 347 (1972).


\end{thebibliography}
\end{document}